\documentclass[preprint,aps,floatfix,showpacs]{revtex4}
\usepackage{graphicx}

\newcommand{\beq}{\begin{equation}}
\newcommand{\eeq}{\end{equation}}
\newcommand{\bea}{\begin{eqnarray}}
\newcommand{\eea}{\end{eqnarray}}
\begin{document}
\preprint{\vbox{\hbox{ JLAB-THY-09-1020} }}
\title{\phantom{x}
\vspace{-0.5cm}  Analysis of {\bf 56}-plet Positive Parity  Baryon Decays 
 in the ${\mathbf 1}{\mathbf /}{\rm \mathbf N_{\mathbf c}}$ Expansion}

\author{
J. L. Goity $^{a,b,f}$ \thanks{e-mail: goity@jlab.org},
 C.  Jayalath $^{a}$  \thanks{e-mail:jayalath@jlab.org}, 
N. N. Scoccola $^{c,d,e}$
\thanks{e-mail: scoccola@tandar.cnea.gov.ar}}

\affiliation{
$^a$ Department of Physics, Hampton University, Hampton, VA 23668, USA. 
$^b$  Thomas Jefferson National Accelerator Facility, Newport News, VA 23606, USA. 
$^c$ Physics Department, Comisi\'on Nacional de Energ\'{\i}a
At\'omica,\\   1429  Buenos Aires, Argentina.
$^d$ CONICET, Rivadavia 1917, (1033) Buenos Aires, Argentina.
$^e$ Universidad Favaloro, Sol{\'\i}s 453,  1078 Buenos Aires, Argentina.
$^f$ Instituto Balseiro,  Centro At\'omico Bariloche, 
8400 S.C. de Bariloche, Argentina.}
\begin{abstract}   
The partial decay widths of positive parity baryons belonging to {\bf 56}-plets of $SU(6)$ are analyzed in the framework of the $1/N_c$ expansion. The channels considered are those with emission of a single  $\pi$,  $K$ or $\bar{K}$ mesons, and the analysis is carried out to subleading order in $1/N_c$ and to first order in $SU(3)$ symmetry breaking.  The results for the multiplet $[{\bf 56},0^+]$, to which the Roper resonance belongs, indicate a poor description at leading order, requiring important  next to leading order corrections. For the multiplet $[{\bf 56},2^+]$, the P-wave decays in the non-strange sector are well described at leading order,  while the F-wave decays require the next to leading order corrections, which turn out to be  of natural magnitude.  $SU(3)$ breaking effects are poorly determined, because only few decays with $K$ meson in final state are established, and their widths are not known  with sufficient accuracy.
\end{abstract}

\pacs{14.20.Gk, 12.39.Jh, 11.15.Pg}

\maketitle

\section{Introduction}
One of the most important objectives in hadronic physics is the increasingly accurate determination of the properties of baryon resonances, the search for predicted and yet unobserved resonances, and the theoretical description and understanding of the resonances'  observables. The study of baryons, which is complementary to that of mesons,   plays indeed an important role in exposing   non-perturbative aspects of QCD,  such as the ordering of states into approximately linear Regge trajectories, the  various strong and electromagnetic transition observables,  and the remarkable identification of spin-flavor multiplets   in the known spectrum of baryons. This  latter property was identified  back in the 1960s \cite{Close}, and in QCD it can be explained in terms of the $1/N_c$ expansion:  in the large $N_c$ limit, baryons must fill multiplets of the contracted spin-flavor symmetry group $SU^{\rm c}(6)$ \cite{GervaisSakita,DashenManohar}. In the real world with $N_c=3$,  we do have evidence of an $O(3)\times SU(6)$ multiplet structure, in particular thanks to the well established ground state octet and decouplet states which are accommodated in
a $[{\bf 56},\ell^P=0^+]$, and the convincingly established $[{\bf 56},2^+]$ and $[{\bf 70},1^-]$  multiplets    in the second resonance region, which, although incomplete, have enough known states   for a good identification of them.  For other multiplets,  the identification needs further scrutiny, based on mass formulas in particular. One such a multiplet is the one containing the Roper resonance
$N(1440)$, which is assumed  to be a    $[{\bf 56},0^+]$. Only  a few of the resonances  which would  be assigned to that multiplet  are    established. 

While in the large $N_c$ limit there must be a contracted $SU^{\rm c}(6)$ symmetry in the baryon sector,   its breaking  happening at ${\cal{O}}(1/N_c)$,  there is no  justification of principle for the observed approximate $O(3)$ symmetry. Within a given  $O(3)\times SU(6)$ multiplet,  the $O(3)$ symmetry is recovered in the large $N_c$ limit if the multiplet corresponds to the symmetric spin-flavor representation, but it is broken at  ${\cal{O}}(N_c^0)$ for a mixed-symmetry multiplet. Thus for the states we discuss in this paper, and provided we neglect configuration mixings \cite{Mixing}, i.e. mixings with, for instance, a mixed-symmetry multiplet, the  breaking of the $O(3)$ symmetry is a sub-leading effect in $1/N_c$. 

  Based on the  $O(3)\times SU(6)$ symmetry scheme, it is possible to implement a $1/N_c$ expansion in baryons  as an effective theory built in terms of effective operators associated with the observable to be analyzed. This framework has been utilized in the analysis of baryon masses \cite{masses,GSmasses}, strong decays  \cite{decays,DecaysSU6,GSdecays70,GSdecays}, magnetic moments \cite{magmom}, electromagnetic helicity amplitudes \cite{helampl}, etc. From these works, it transpires that the ordering of effects according to the $1/N_c$ power counting is remarkably well manifested for the most observables, that have  been studied, with some  exception in  the case of the $[{\bf 56},0^+]$ Roper multiplet,  which is  part of the present analysis.

This work extends the analysis of the strong decays of positive parity baryons studied within $SU(4)$ \cite{GSdecays} to  the ${\bf 56}$-plet of $SU(6)$. The analysis is carried out to subleading  ${\cal{O}}(1/N_c)$ and to first order in $SU(3)$ symmetry breaking, and  the framework follows a similar  implementation as in the case of  decays in $SU(4)$ \cite{GSdecays}.  An earlier analysis of the decays of the  $[{\bf 56},0^+]$ baryons in the $1/N_c$ expansion was  made in Ref. \cite{DecaysSU6}. The present work presents  a  full fledged analysis to the orders mentioned for that multiplet as well as for the   $[{\bf 56},2^+]$. 

The inputs utilized for the analysis are the partial decay widths as given in terms of fractions of the Breit-Wigner widths by the Particle Data Group \cite{PDG}. Most of these inputs have changed only slightly over the last few editions of the particle listings, and we refer to the latest edition for references to the most recent partial wave analyses.

The work is organized as follows: in section II, we outline the implementation of the $1/N_c$ expansion for the decays; in section III, we give the results for the analyses of the $[{\bf 56},0^+]$ and the $[{\bf 56},2^+]$ baryon decays,  followed by the conclusions  in section IV.  For completeness, an appendix provides details on the $SU(3)$ isoscalar factors   needed  in the calculations.
 
\section{Theoretical background}
\subsection{Baryon states in $O(3)\times SU(6)$ }

The excited baryon states in the large $N_c$ limit must fill towers of   states, which correspond to irreducible representations of a contracted spin-flavor symmetry  group $SU^{\rm c}(2 N_f)$. The states of angular momentum of order unity  in these towers have mass splittings  ${\cal{O}}(1/N_c)$.  As already mentioned, in the real world with $N_c=3$,  it is observed that there is in addition an approximate $O(3)$ symmetry, which  does not follow from the $1/N_c$ expansion but  is rather from phenomenological observation.  
This symmetry is most clearly  displayed in the spectrum of the known excited baryons. Thus, the classification of the excited baryons in terms of the symmetry group $O(3) \times  SU(2 N_f)$ is the proper approach. In large $N_c$, the contracted symmetry group will emerge  as a subgroup of that larger group \cite{PirjolSchat}.   
 
For three light flavors, we need the group $SU(6)$, which
has thirty five  generators, namely  $\{S_i,T_a,G_{ia}\}$, with $i=1,2,3$ and $a=1,\cdots,8$,  where the first
three are the generators of the spin $SU(2)$, the second eight are the generators of flavor $SU(3)$, and the last
twenty four can be identified as an octet of axial-vector  currents in the limit of zero momentum transfer. The
algebra of $SU(6)$ has the following commutation relations that fix the normalizations of the generators:
\bea
\left[S_i,S_j\right] = i\,\epsilon_{ijk} S_k\; , &&
\left[T_a,T_b\right]  = i\, f_{abc} T_c \; , \nonumber \\
\left[S_i,G_{ja}\right]  =  i\,\epsilon_{ijk}G_{ka}\; , &&
\left[T_a,G_{ib}\right]  =  i\,f_{abc}G_{ic}\; ,\nonumber \\
\left[G_{ia},G_{jb}\right] =  i\, \delta_{ij} f_{abc} T_c & \! +\! &
i\, \epsilon_{ijk}(\delta_{ab} S_k+d_{abc} G_{ck})\; ,
\eea
where  $d_{abc}$ and $f_{abc}$ are the  $SU(3)$ symmetric and antisymmetric  invariant tensors, respectively.

  The states of interest in this work belong to the totally symmetric irreducible representation {\bf S},  which is given by a Young tableau  consisting  of a
single row of  $N_c$ boxes. These states correspond at $N_c=3$ to the ${\bf 56}$-plet of $SU(6)$. Another important representation is the   mixed symmetric   {\bf MS},  which  consists of
a row with $N_c-1$ boxes and a second row with a single box,  that  for $N_c=3$  corresponds to the ${\bf 70}$-plet of $SU(6)$. We denote, by $\ell$, the $O(3)$ quantum number of the states.  The states belonging to the $[{\bf 56},\ell]$ multiplet    are  then  given by
\begin{equation}
 |(\ell,S)J J_3;  R=(p,q), Y, I  I_3 >\!\!\!_{_{\rm \bf S}}
= \sum_{m,S_3}
 \langle   \ell m,\, S   S_3\mid   J  J_3\rangle
    |S S_3  ;  R=(p,q), Y,I I_3   >\!\!\!_{_{\rm \bf S}}\mid \ell\  m \rangle,
\end{equation}
where the label {\bf S} indicates that the state belongs to  the symmetric spin-flavor representation, $S$ is the  spin quantum number  associated with the spin subgroup of $SU(6)$,  $J$ is the total angular momentum of the baryon,     $R=(p,q)$ indicates the  $SU(3)$ irreducible representation given in terms of the usual labeling of a Young tableu with $p+2 q=N_c$,  and Y and I are respectively the hypercharge and the isospin.


 From the decomposition of the {\bf  S}  representation of $SU(6)$
as a sum of direct products of  irreducible representations of $SU(2)\otimes SU(3)$,  it results that
 $p=2 S$. This latter relation is a consequence of the fact that,  for the {\bf S}  representations, the two factors in the direct
products involved in the decomposition  have the same Young tableau $(p,q)$. The relation then follows  from the  relation $p=2 S$ in  $SU(2)$.   The   $p=2 S$ relation is a generalization of the  $I=S$ relation for the {\bf S} representations of $SU(4)$.

The correspondence of multiplets between generic $N_c>3$ and $N_c=3$ is as follows: i) $(p=1, q=\frac{N_c-1}{2})\to ( {\bf 8},\;S=\frac{1}{2})$ and ii) $(p=3, q=\frac{N_c-3}{2})\to ({\bf 10},\;S=\frac{3}{2})$. Clearly, all  this holds for odd $N_c$.

For $\ell=0$, which also includes the ground state (GS) baryons, we have the $J=S=1/2$ octet and the $J=S=3/2$ decuplet. For $\ell=2$, we have octets with
$J=3/2$, $5/2$ and decuplets with $J=1/2$ through $7/2$.

 \subsection{ $O(3)\times SU(6)$ framework for decays}

In the following,  the framework for implementing the calculation of decays via emission of a single pseudoscalar meson is described. 
We first give  the partial decay widths in terms of reduced matrix elements of the effective baryonic operators,  which describe the transition between the initial excited baryon and the final ground state baryon. If we   altogether neglect $SU(3)$ symmetry breaking, the partial widths could be expressed solely in terms of $SU(3)$ reduced matrix elements of the effective operators. However, $SU(3)$ symmetry breaking is noticeable in the decays, and must be included. This is done  with  effective operators, generated by the octet component of the quark masses (i.e., by the term $\frac{1}{2 \sqrt{3}}\,(m_u+m_d-2 m_s )\lambda_8$ in the quark mass matrix), and thus the partial widths are rather given in terms of reduced matrix elements of the isospin $SU(2)$. We start from the general expression for a partial decay width of an initial excited baryon state $\mid (\ell, S^*) J^*, J^*_3; R^*,Y^*,I^*,I^*_3\rangle $  to a final state  $\mid   S,S_3; R ,Y ,I ,I_3\rangle $ emitting a pseudoscalar meson in the state $\mid \tilde{\ell}, \tilde{\ell}_3;\tilde{R},\tilde{Y},\tilde{I},\tilde{I}_3\rangle$. The effective operator must carry the same quantum numbers of the meson. The partial decay width is then given by:
\bea
\Gamma & =&\frac{k}{8 \pi^2}\;\frac{M_B}{M_{B^*}}\;\frac{1}{(\hat{I^*} \hat{J^*})^2}\nonumber\\
&\times&\sum_{J^*_3,I^*_3,S_3,I_3,\tilde{\ell}_3,\tilde{I}_3}  \arrowvert
\langle S,S_3; R ,Y ,I ,I_3\mid B^{[ \tilde{\ell},  \tilde{R}]}_{[\tilde{\ell}_3 ,\tilde{Y},\tilde{I},\tilde{I}_3]}\mid (\ell, S^*) J^*, J^*_3; R^*,Y^*,I^*,I^*_3\rangle \arrowvert^2,
\eea
where $M_{B^*}$ and $M_{B}$ are the masses of the excited and final baryon respectively,  $k$ is the meson momentum, and the notation $\hat{j}\equiv \sqrt{2 j+1}$ is used. The baryonic transition operator is denoted by $B^{[ \tilde{\ell},  \tilde{R}]}_{[\tilde{\ell}_3 ,\tilde{Y},\tilde{I},\tilde{I}_3]}$,  where the upper labels  display  the
angular momentum of the operator and its $SU(3)$ representation, and the lower labels  the corresponding projections.
The operator is built as a linear combination in a basis of effective operators, which is ordered in powers of $1/N_c$:
\beq
B^{[  \tilde{\ell},\tilde{R}]}=
\left(\frac{k}{\Lambda}\right)^{\tilde{\ell}}
\sum_n
C_n^{[ \tilde{\ell},  \tilde{R}]}(k)\;B_n^{[ \tilde{\ell},  \tilde{R}]},
\eeq
where the  $B_n$ are the operators in the basis, and the coefficients $C_n$ encode the dynamics of the decay amplitudes.  In this work, they are determined by fitting to the empirical partial decay widths. A centrifugal factor is included,  which is expected to carry the chief momentum dependence of the transition amplitude, with the arbitrary scale $\Lambda$ to be chosen to be $200$ MeV in what follows.

The effective operators in the basis can be expressed in terms of spin-flavor operators in the following general form:
\beq
B^{[\tilde{\ell},\tilde{R}]}_{n~[\tilde{\ell}_3,\alpha]}=
\sum_{m,j_3}\langle \ell m,j_n j_3 \mid \tilde{\ell} \tilde{\ell}_3\rangle\;\xi^\ell_m \;{\cal{G}}^{[j_n,\tilde{R}]}_{n~[j_{3},\alpha]},
\eeq
where  $\xi^\ell$ is the tensor operator, which gives the transition between the   $O(3)$ state of the excited baryon and the ground state baryon,   and ${\cal{G}}_n^{[j,\tilde{R}]} $ is a spin-flavor operator.  Without any loss of generality,  $\xi^\ell$ is normalized such that   $\langle 0 \mid \xi^\ell_{m'}\mid \ell m\rangle=
(-1)^{\ell-m}\delta_{m\;-m'}$. 
 The spin-flavor operator, which is a tensor  with angular momentum $j_n$,  can be built as tensor products of the $SU(6)$ generators.

We can express the partial widths in terms of the RMEs of the basis operators as follows:
\bea
\Gamma&=& \frac{k}{8\pi^2}
\left(\frac{k}{\Lambda}\right)^{2\tilde{\ell}}\frac{M_B}{M^*_B}
\;\frac{\hat{I}^2}{(\hat{I^*} \hat{J^*})^2}\nonumber\\
&\times& 
\arrowvert \sum_n C_n^{[ \tilde{\ell},  \tilde{R}]}(k)
 \;{\bf B}_n(\{S,R,Y,I\},\{(\ell,S^*)J^*,R^*,Y^*,I^*\},\{\tilde{\ell},\tilde{R},\tilde{Y},\tilde{I}\} )    \arrowvert^2,
\eea
where  the  ${\bf B}_n$  are the RMEs. These    can also be expressed in terms of RMEs of spin-flavor operators. For operators $B_n$, which do not involve $SU(3)$ symmetry breaking, one obtains:
\bea
& &{\bf B}_n(\{S,R,Y,I\},\{(\ell,S^*)J^*,R^*,Y^*,I^*\},\{\tilde{\ell},\tilde{R},\tilde{Y},\tilde{I}\}) \nonumber\\
&=&(-1)^{j_n+J^*+\ell+S}\;\frac{\hat{J^*}\;\hat{\tilde{\ell}}}{\sqrt{{\rm dim}\; R}} 
\left\{
 \begin{array}{ccc}
     J^*   & S^*     & \ell   \\
     j_n    & \tilde{\ell}     & S
\end{array} \right\} 
\; \sum_\gamma
\left(\begin{array}{cc}
     R^*   & \tilde{R}        \\
     Y^*~I^*     & \tilde{Y} ~\tilde{I}
\end{array} \right\Arrowvert   \left.  \begin{array}{c}
      R         \\
     Y~~ I
\end{array} \right)_\gamma \nonumber\\
&\times& \langle S,R\| {\cal{G}}_n^{[j_n,\tilde{R}]}\|S^*,R^*\rangle_\gamma,
\eea
where,  with obvious notation, there  appear  a $SU(2)$ 6-j symbol,   $SU(3)$ isoscalar
factors,  and the reduced matrix element of the  corresponding spin-flavor operator.
$\gamma$ labels the possible multiplicities for coupling  the product of representations
 $R^*\otimes \tilde{R}$ to $R$ in $SU(3)$. Throughout,  the $SU(2)$ conventions are those
 of Edmonds \cite{Edmonds},  and the $SU(3)$ conventions are  those established in the
 article by Hecht \cite{Hecht}.

 In the case of $SU(3)$ symmetry breaking operators, we proceed as follows. The symmetry breaking is due to the mass difference between the $s$ quark and  the  $u$ and $d$ quarks. To first order in the quark masses,  this symmetry breaking is implemented at the level of the spin-flavor operators according to:
 \beq
 {\cal{G}}_{n,\gamma_n}^{[j_n,\tilde{R}]}=\frac{1}{\Lambda}\,[{\cal{M}}^8_q \; {\cal{G}}_n^{[j_n, R_n]}]_{\gamma_n}^{\tilde{R}},
 \eeq
where  $ {\cal{M}}^8_q$ is   the octet component of the quark mass matrix, $\Lambda$ an arbitrary scale to render operators dimensionless,  and $\gamma_n$ indicates the  particular    coupling ${\bf 8}\otimes R_n$ to $\tilde{R}$.  For the case of interest here, $\tilde{R}={\bf 8}$, and therefore   $R_n$ can be {\bf 1}, {\bf 8}, {\bf 10}, ${\bf \bar{10}}$ or {\bf 27}. These possibilities give  rise to a large proliferation of $SU(3)$ breaking operators at 2-body level.  Fortunately, for the   decays  of 56-plet baryons, those 2-body operators are higher order in $1/N_c$, and thus contribute corrections to the leading order decay amplitudes of order $(m_s-m_{u,d})/N_c$, which are  beyond the accuracy of the present  analysis. This is quite different in the case of the {\bf 70}-plet, as it will be discussed elsewhere.
The RME of an $SU(3)$ breaking operator will then be given by:
\bea
&&{\bf B}_n(\{S,R,Y,I\},\{(\ell,S^*)J^*,R^*,Y^*,I^*\},
\{\tilde{\ell},\tilde{R},\tilde{Y},\tilde{I}\})
\nonumber\\
&=& (-1)^{j_n+J^*+\ell+S}\;\frac{\hat{J^*}\;\hat{\tilde{\ell}}}{\sqrt{{\rm dim}\; R}}
\left\{
\begin{array}{ccc}
     J^*   & S^*     & \ell   \\
     j_n     & \tilde{\ell}     & S
\end{array} \right\}
\left( \begin{array}{cc}
     8   &  R_n        \\
     0~0     &  \tilde{Y}~ \tilde{I}
\end{array}  \right\Arrowvert \left. \begin{array}{c}
      \tilde{R}         \\
    \tilde{ Y}~ \tilde{I}
\end{array}\right)_{\gamma_n}\sum_{\gamma}
\left(\begin{array}{cc}
     R^*   &  R_n        \\
     Y^*~I^*     &  \tilde{Y} ~  \tilde{I}
\end{array}  \right\Arrowvert \left. \begin{array}{c}
      R         \\
     Y~~ I
\end{array} \right)_{\gamma}\nonumber\\
&\times& \langle S,R\| {\cal{G}}_n^{[j_n, R_n]}\|S^*,R^*\rangle_{\gamma}\;\; ,
\eea
where $\gamma_n$ indicates the $SU(3)$ recoupling corresponding to the operator ${\rm \bf B}_n$.
With the definition of effective operators used in this work, all  coefficients
$C_n^{[\ell, \tilde{R}]}(k)$ in Eqn. (4) are of zeroth order in $N_c$. In addition,
we will normalize the operators in such a way that the {\it natural} size of all coefficients
would be the same.
The leading order of the
decay amplitude is in fact $N_c^0$ \cite{Mixing}.
At this point, it is important to comment on the  momentum dependence of
the coefficients. The  spin-flavor breakings  in the masses, of both excited
and ground state baryons,  give rise to different values of the momenta $k$.
In the  56-plets the mass splittings   are   however ${\cal{O}}(1/N_c)$ or order $m_s-m_{u,d}$, therefore, those effects on $k$ are
 taken into account  automatically in the expansion, we are performing. Thus,   we can ignore any momentum dependence of the coefficients $C_n^{[\ell, \tilde{R}]}$ as such  effects are  absorbed into the operators.

\subsection{Operator basis }

The construction of a basis of spin-flavor operators follows
similar lines as in previous works on baryon decays \cite{GSdecays}.  The main difference between the $SU(4)$ and $SU(6)$
cases is that, in the first case, the matrix elements of a given spin-flavor operator are all of the same order in $1/N_c$, while in the latter case, they are not. For instance,  the emission of $K$ mesons is suppressed by a factor $1/\sqrt{N_c}$ with respect to the emission of pions. This is due to the fact that in the $1/N_c$ counting baryons are considered to have strangeness ${\cal{O}}(N_c^0)$. Note that  this does not represent any $SU(3)$ symmetry breaking.  Due to this issue, we classify operators, which are $SU(3)$ preserving according to the order in $1/N_c$, at which they contribute in pion emission. The order, at which  an $n$-body operator contributes  to the transition amplitude, is  given by $\nu=n-\kappa$,  where  $\kappa$ is its degree of coherence factor determined by the number of coherent generators, that appear in the product building the   operator. Details on the derivation of this power counting can be found in \cite{Mixing}.

In the decays of the positive parity baryons, we have odd partial waves, and we analyze the $P$ and $F$ waves. The $SU(3)$ preserving operators for the emission of the pseudoscalar octet are as follows. There is only one 1-body operator, namely $G_{ia}/N_c$. This operator is leading order, and gives contributions to the decay widths into pions at ${\cal{O}}(N_c^0)$. Next, we have 2-body operators, which are built by multiplying a pair of generators such that one can couple them to $j=1$ or $2$ and to $R={\bf 8}$. One can construct six such products, which upon using the $SU(6)$ reduction formulas \cite{JDM} and keeping only up to operators
 ${\cal{O}}(1/N_c)$, only two operators are left, namely $1/N_c^2\;\{S,G\}^{[j=1,{\bf 8}]}$ and  $1/N_c^2\;\{S,G\}^{[j=2,{\bf 8}]}$.  One can show that  3-body operators  contribute to amplitudes at ${\cal{O}}  (1/N_c^2)$,  which is beyond the order of this work.

To the order we are working,   $SU(3)$ breaking  is described by  two   1-body operators,  namely  $f_{8ab} G_{ib}/N_c$ and $d_{8ab} G_{ib}/N_c$. The first operator does not contribute to decays involving a $\pi$ meson, and the second one    can be  redefined   by adding to it  a non-breaking piece in such a way that it does not contribute to matrix elements involving a $\pi$ meson. Due to the small number of empirically known decay channels into K-mesons, we will not be able to fit the effects of both operators. For this reason, only the operator  $d_{8ab} G_{ib}/N_c$ will be utilized.

 The $1/N_c$ counting for the reduced matrix elements of operators is finally given as follows: i) for $SU(3)$ preserving LO operators, the amplitudes $B^*\to \pi B$ are  ${\cal{O}}(N_c^0)$ and $B^*\to K B $ or $B^*\to  \bar{K} B$ are ${\cal{O}}(N_c^{-1/2})$; NLO amplitudes are simply an extra factor $1/N_c$ in all cases. ii) for the $SU(3)$ breaking operator, once re-defined to have vanishing contributions to amplitudes with pion emission, it  contributes  at  ${\cal{O}}(m_s\;N_c^{-1/2})$.

Table I summarizes the set of spin-flavor operators relevant to the decays of the ${\bf 56}$-plet baryons as needed in this work. 
 \begin{center}
\begin{table}[h]
\caption{ Spin-flavor operators }
\vspace{.5cm}
 \begin{tabular}{cc}
   \hline\hline
   Operator  & Order \\ [1.mm] \hline
    ${\cal{G}}_1\equiv\frac{1}{N_c}\; G$&${\cal{O}}(N_c^0)$ \\
    ${\cal{G}}_2\equiv\frac{1}{N_c^2}\; \{S,G\}^{[j=1]} $&${\cal{O}}(1/N_c)$ \\
     ${\cal{G}}_3\equiv\frac{1}{N_c^2}\; \{S,G\}^{[j=2]} $&${\cal{O}}(1/N_c)$ \\
      ${\cal{G}}_{SB}\equiv\frac{1}{N_c}\;(d_{8ab}-\delta_{ab}/\sqrt{3})\; G_{ib}$& $~~~~{\cal{O}}((m_s-m_{u,d})/\sqrt{N_c})$ \\ [2.mm]
   \hline\hline
 \end{tabular}
 \end{table}
 \end{center}

The   RMEs  of the spin-flavor operators between the ${\bf 56}$-plets    can be expressed  in terms of the $SU(3)$ reduced matrix elements of $G$ and of $\{S,G\}$.  
The RMEs of $G$ are given in Table II, and the ones of   $\{S,G\}^{[j]}$ are related to those of $G$ by the following formula:
\bea
\langle R'  \|  \{S,G\}^{[j,{\bf 8}]} \| R \rangle_{\gamma}&=& (-1)^{S+S'}\; \hat{j}\; \langle R'  \|   G \| R \rangle_{\gamma}\nonumber \\
\times\left((-1)^j\,\hat{S'} \sqrt{S'(S'+1)}
\left\{
\begin{array}{ccc}
     S' & S' & 1   \\
     1  & j & S
\end{array} \right\}\right.&+&\left. \hat{S}\sqrt{S(S+1)}
\left\{
\begin{array}{ccc}
     S & S & 1   \\
     1  & j & S'
\end{array} \right\}\right)\; ,
\eea
where,   as mentioned earlier, $S$ and $S'$ in the ${\bf 56}$-plet are  determined by the respective $SU(3)$ representation $R$ and $R'$. In this work, we only need the cases $j=1,2$.
 \begin{center}
\begin{table}[h]
\parbox{3in}{\caption{$SU(3)$     RMEs  of the spin-flavor operator $G$ between  $SU(3)$  multiplets in the $SU(6)$ ${\bf 56}$-plet.    $\langle {\bf 10}  \| G \| {\bf 8}\rangle=\langle {\bf 8}  \| G \| {\bf 10}\rangle$, and  $\langle {\bf 10} \|  G \|  {\bf 10}\rangle_{\gamma=2} =0$ when $N_c=3$.}}\\
\vspace{.5cm}
\begin{tabular}{ l  }
\hline\hline
~~~~~~~~~~~~~~~~~~~~~~~~~~~~~~~~RME  \\ [1.mm] \hline \\ [-3.mm]
    $\langle {\bf 8}  \| G \| {\bf 8} \rangle_{\gamma=1} = \sqrt{\frac{(N_c+1)(N_c+5)}{2}}$   \\
    $\langle {\bf 8}  \| G \| {\bf 8} \rangle_{\gamma=2} = -\sqrt{\frac{(N_c^2-1)(N_c+5)(N_c+7)}{32}}$    \\
    $\langle {\bf 8}  \| G \| {\bf 10}\rangle   = -\frac{1}{4}\,\sqrt{(N_c^2-1)(N_c+5)(N_c+7)}$     \\
    $\langle {\bf 10} \| G \| {\bf 10}\rangle_{\gamma=1}  =  (N_c+3)\;\sqrt{\frac{10(N_c-1)(N_c+7)}{(45+N_c(N_c+6))}}$   \\
    $\langle {\bf 10} \|  G \|  {\bf 10}\rangle_{\gamma=2} = -\frac{1}{\sqrt{8}}\;\sqrt{\frac{(N_c^2-1) (N_c-3) (N_c+5)(N_c+7)(N_c+9)}{(45+N_c(N_c+6))}}$    \\ [2.mm]
\hline\hline
\end{tabular}
\end{table}
\end{center}
Notice that the $1/N_c$ power counting is not given solely by the reduced matrix elements of the baryonic operator, but also involves  the $N_c$ dependence of the isoscalar factors. For this reason,  we give in an appendix  the expressions for general $N_c$ of those isoscalar factors as needed in the present calculations.

\section{Results}
In this section, we present and analyze the fits to partial widths. In general, the LO fits correspond to the results one would obtain from coupling the pseudoscalar meson to the excited quark, similar to the framework of the chiral quark model, and also give results, which are similar to those in quark models in general \cite{RobertsCapstick}. Our analysis below shows the shortcomings of the LO approximation, and the need for the NLO contributions, which are given by 2-body effects in the decay amplitudes.

\subsection{Results for the $[{\bf 56},0^+]$ decays}
The basis of  operators  in this case contains  the LO ${\cal{G}}_1$ operator, the 2-body  NLO  ${\cal{G}}_2$ operator and the $SU(3)$ breaking ${\cal{G}}_{SB}$ operator.     We normalize these operators according to  ${\cal{G}}_n\to \alpha_n  {\cal{G}}_n$, with
$\alpha_1=6/5$, $\alpha_2=1/\sqrt{2}$, $\alpha_{SB}=2 \sqrt{3}/5$,  so that the  corresponding matrix elements have natural magnitude.
 In this multiplet, the experimentally established states to be used in the analysis along with their respective star rating by the Particle Data Group are:  the Roper  resonance $N(1440)$ ($\ast\!\ast\!\ast\ast$), the $\Lambda(1600)$ ($\ast\!\ast\!\ast$) and the $\Sigma(1660)$  ($\ast\!\ast\!\ast$), all of them in the {\bf 8},  and only the $\Delta(1600)$  ($\ast\!\ast\!\ast$)  in the {\bf 10}. 
 
 The results of the fits are shown in Table III.

\begin{center}
\begin{table}[h]
\parbox{5in}{\caption{Fit parameters and partial widths in MeV for P-wave decays of the $[56,0^+]$ baryons.
 The theoretical errors of the results from the  NLO fit are indicated explicitly; similar errors result for the NLO* fit.}}\\
\vspace{.5cm}
\begin{tabular}{c c cc c ccccc}
\hline \hline
     &&  $\chi^2_{dof}$ & dof && $C_1$ && $C_2$ &&  $B_1$ \\ \hline
LO       &&  2.2 & 6 &&  8.8(1.3)&& $-$ && $-$  \\
${\rm LO^*}$   &&  1.7 & 5 &&  6.7(0.8)                      && $-$ && 4.2(3.2)\\
NLO  && 0.8   & 4 &&  8.1(0.9)&&  -8.0(2.6)&&2.7(4.6)
\\
${\rm NLO^*}$  && 0.9   & 4 && 8.7(1.0) &&  -9.3(4.3) &&  8.2(3.8)
\\[1mm] \hline \hline
\\
\end{tabular}
\begin{tabular}{cccccccccccccccccc}
\hline \hline
  &  &\multicolumn{2}{c}{$N_\frac12 (1440)$}&
     &\multicolumn{3}{c}{$\Lambda_\frac12 (1600)$}&
     &\multicolumn{4}{c}{$\Sigma_\frac12 (1660)$}&
     &\multicolumn{2}{c}{$\Delta_\frac32(1600)$}
\\[1mm]
\cline{3-4}\cline{6-8}\cline{10-13}\cline{15-16}
  &  &  $\pi N$ & $\pi \Delta$ &
    & $\bar K N$ & $\pi \Sigma$ & $\pi \Sigma^*$ &
    & $\bar K N$ & $\pi \Lambda$ & $\pi \Sigma$ & $\pi \Sigma^*$ &
    & $\pi N$ & $\pi \Delta$ \\
\cline{3-4}\cline{6-8}\cline{10-13}\cline{15-16}
LO  &  &   90     &  12          &
&    0       &  73     & 24    &
&     0   &   41   & 58  & 6   &
&  93    &  55
\\
${\rm LO^*}$  &  &   148     &  13          &
&    36       &  57     & 19    &
&     1.6   &   32   & 61  & 6.5   &
&  97    &  90
\\
NLO &   &   214    &   65         &
&   38       &  111    &  95   &
&   5     &   63   & 88  & 25  &
&   56    &   131
\vspace{-0.25cm}\\
 &   &   (49)    &   (17)         &
&   (28)       &  (25)    &  (27)   &
&  (4)     &   (14)   & (20)  & (7)  &
&   (44)    &   (30)
\\
${\rm NLO^*}$ &   &   250    &   45         &
&   36       &  97   &  66  &
&   1.6     &   55   & 103  & 23  &
&   51    &   153
\\
Exp  &  &  $211(88)$ & $81(35)$   &
& $34(25)$ & $53(51)$  &   $-$    &
&  24(20) &    $-$    &  $-$   &  $-$ &
&  $61(32)$ & $193(76)$
\\[1mm] \hline \hline
\end{tabular}
\end{table}
\end{center}

In all the leading order fits considered here denoted as LO, the decays with $K$ or $\bar{K}$ mesons in final state are not included, because their widths start at ${\cal{O}}(1/N_c)$.  The LO fit involves only one operator, and has been carried out by assigning error bars to the input partial widths, obtained from the PDG, which are 30\% or the experimental value if it is larger than 30\%. This is to test whether  or not the LO is consistent. 
In the case of the $[{\bf 56},0^+]$  decays,  it  turns out that the LO fit is not consistent.  This was well known from previous work in $SU(4)$ \cite{GSdecays},  where the non-strange channels were shown to be   poorly described.  The problem is best
illustrated by the two ratios, which are parameter-free  at LO :  i) $\Gamma(N(1440)\to\pi N)/\Gamma(N(1440)\to\pi \Delta)=$7.5 (Th) vs $2.6\pm2.1$ (Exp), and ii)  $\Gamma(\Delta(1600)\to\pi N)/\Gamma(\Delta(1600)\to\pi \Delta)=$1.7 (Th) vs $0.3\pm 0.2$ (Exp). Thus, spin-flavor symmetry is badly broken in these decays; this is the most notable inconsistency of the $1/N_c$ expansion, and points to the particular nature of the Roper multiplet. At LO, the decay rate $\Lambda(1600)\to \pi \Sigma$ is remarkably close to the observed one, but not so well described at NLO.  As in the other fits in this work, we include a fit denoted by LO*, in which we do not expand the
matrix elements of the LO operator, and also include the $SU(3)$ breaking operator.  The reason for doing this is that   some of the reduced matrix elements and isoscalar factors are poorly represented by the first term in the expansion. This fit gives significant improvement over the LO one, shows that the $SU(3)$ breaking operator is nearly irrelevant, and it further exposes the problematic decay channels, which are the $\pi \Delta$ channels. 
Therefore, the NLO corrections  must be very important, and they are provided by the single  2-body  operator. The inclusion of these NLO corrections  leads to a consistent fit. One issue is that the   $\Lambda(1600)\to \pi \Sigma$ channel is not as well described as at LO. The main NLO effect is to enhance the $N(1440)\to \pi \Delta$ width  to become consistent with the experimental one. One point to be noted is that the coefficient of the NLO operator is of natural size,  although this gives  NLO effects in some of the widths, which are large.  In the NLO*, we do not expand the matrix elements. The main effect is to change the coefficients $C_2$ and $B_1$, and the reason is that the NLO operator does not contribute to the decays into $K$ mesons in the NLO fit, while such contributions do occur in NLO*. 
 
For the $SU(3)$ symmetry breaking we find, for all the fits in this work,  two consistent solutions.  In all cases,  one of the solutions has a coefficient, which is unnaturally large, and we assume that such a solution is unacceptable. The only way to confirm  this assumption   is by determining empirically other widths sensitive to the symmetry breaking.  In the case of the $[{\bf 56},0^+]$,  the role of the $SU(3)$ breaking operator cannot be established:  it is  irrelevant in the NLO fit and relevant in NLO* fit.   The reason for this ambiguity is that only two decay widths are involved, and both have large error bars.
Finally, the fits permit us to give rough predictions for the unknown channels.   
 
\subsection{Results for the  $[{\bf 56},2^+]$ decays}
The basis of  operators  in this case involves   the LO ${\cal{G}}_1$ operator, the NLO  ${\cal{G}}_2$  and ${\cal{G}}_3$ 2-body operators  and the $SU(3)$ breaking ${\cal{G}}_{SB}$ operator.     In this case, the normalization factors are taken to be as follows: i) for the P-wave decays 
$\alpha_1=4/\sqrt{7}$, $\alpha_2=1$, $\alpha_3=\sqrt{3}/2$,  $\alpha_{SB}=4/\sqrt{21}$,  and
 ii) for F-wave decays  $\alpha_1=3/\sqrt{8}$, $\alpha_2=1/2$, $\alpha_3=\sqrt{3/10}$,  $\alpha_{SB}=\sqrt{3/8}$.

The experimentally established states in the $[{\bf 56},2^+]$ to be used in the analysis are the following ones. In the octets: i) $J=3/2$:  $N(1720)$  ($\ast\!\ast\!\ast\ast$)  and  $\Lambda(1890)$  ($\ast\!\ast\!\ast\ast$), ii) $J=5/2$:  $N(1680)$  ($\ast\!\ast\!\ast\ast$), $\Lambda(1820)$  ($\ast\!\ast\!\ast\ast$) and $\Sigma(1915)$  ($\ast\!\ast\!\ast\ast$). In the decuplets: i) $J=1/2$: $\Delta(1910)$  ($\ast\!\ast\!\ast\ast$), ii) $J=3/2$: $\Delta(1920)$  ($\ast\!\ast\!\ast$),  iii)  $J=5/2$: $\Delta(1905)$  ($\ast\!\ast\!\ast\ast$), iv) $J=7/2$: $\Delta(1950)$  ($\ast\!\ast\!\ast\ast$) and $\Sigma_{10}(2030)$  ($\ast\!\ast\!\ast\ast$).

Let us discuss the results for the P- and F-wave decays separately.

$\bullet$ P-wave decays: All states with $J<7/2$ have a P-wave decay channel. At LO, there is only one operator and the fit  is consistent as previously observed \cite{GSdecays}, giving $\chi^2_{dof}\sim 1.6$.  The main inconsistency appears in the decay $\Lambda(1890)\to\pi\Sigma$, and this problem gets improved but not  solved at NLO. We also show a LO* fit as described earlier. This fit gives further improvement over the LO one, and shows the $SU(3)$ breaking operator to be insignificant. 
Since, at LO*, one already has a very good description, one expects the NLO corrections to be small, 
 which is indeed the case.
 As it occurs for the  $[{\bf 56},0^+]$, the breaking of $SU(3)$  cannot be accurately determined, because there are only two widths with large error bars, which are  affected.
  We note some rearrangement in going from NLO to NLO*, in the first case the breaking being rather large, and in the second one is marginally relevant. Thus, the $1/N_c$ expansion works particularly well in P-wave decays, but the $SU(3)$ breaking cannot be clearly determined.
  
We have included predictions for the decay rates into $\eta$ mesons. Although these widths are suppressed, starting at 
${\cal{O}}(1/N_c^2)$, it is interesting to have an idea of their magnitude. The fact that no such decays for states assigned to {\bf 56}-plets are experimentally observed is a strong indication of the correctness of such assignments.  For {\bf 70}-plet baryons, the $\eta$ modes are not suppressed and start at
${\cal{O}}(N_c^0)$ \cite{GSdecays70}.

\begin{center}
\begin{table}
\parbox{5in}{\caption{Fit parameters and partial widths in MeV for P-wave decays of the $[56,2^+]$ baryons.}}\\
\vspace{.5cm}
\begin{tabular}{c c cc c ccccccc}
\hline \hline
     &&  $\chi^2_{dof}$ & dof &&   $C_1$     && $C_2$       && $C_3$       && $B_1$
\\[1mm]
\hline
LO   &&  1.6   & 6   &&   2.8(0.4) && $-$         && $-$         && $-$
\\
${\rm LO^*}$   &&  0.7    & 5   &&   2.2(0.3)&& $-$         && $-$         &&  -1.2(1.6)\\
NLO  &&  0.9     & 3   &&  1.9(0.2)&& 0.2(1.2)   &&  0.4(2.8)   && -4.2(2.2)
\\
${\rm NLO^*}$  &&  1.2    & 3   &&  2.2(0.3)&& 0.2(1.0)   &&  0.5(2.3) &&  -1.3(1.6)
\\[1mm]
\hline \hline
\\
\end{tabular}
\begin{tabular}{ccc ccccc c ccccc c c c c c cc}
\hline \hline
&&& \multicolumn{5}{c}{$N_{\frac{3}{2}}(1720)$} &&
    \multicolumn{5}{c}{$\Lambda_{\frac{3}{2}}(1890)$}&
   & $N_{\frac{5}{2}}(1680)$ &
   & $\Lambda _{\frac{5}{2}}(1820)$ &
   & \multicolumn{2}{c}{$\Sigma_{\frac{5}{2}}(1915)$ }
\\[1mm]
\cline{4-8}\cline{10-14}\cline{16-16}\cline{18-18}\cline{20-21}
&&& $\pi N$ &  $\eta N$ & $K \Sigma$ & $K \Lambda$  & $\pi \Delta$  &
  &  $\bar{K} N$ &  $\pi \Sigma$  & $\eta \Lambda$ & $K \Xi$  & $\pi \Sigma^*$ &
  & $\pi  \Delta$  &
   & $\pi  \Sigma^*$ &
   & $\bar{K} \Delta$ &  $\pi  \Sigma^*$\\
\cline{4-8}\cline{10-14}\cline{16-16}\cline{18-18}\cline{20-21}
LO
&& &
21 & 0 & 0 & 0 & 1.7 &
  & 0 & 19 & 0 & 0 & 1.8 &
 & 8.1 &
 & 7.7 &
 & 0 & 2.2
\\
${\rm LO^*}$
&& &
36 & 1 & 0.1 & 3 & 2 &
  & 38 & 16 & 4 & 0.3 & 1.6 &
 & 9.2 &
 & 6.6 &
 & 9.6 & 2.5
\\ 
NLO
&& &
 27 & 2.4 & 0.2 & 2.9 & 2.5 &
   & 38 & 16 & 9.8 & 1 & 1.8&
 & 10 &
 & 6.1 &
 & 7.2 & 1.8
\vspace{-0.25cm}\\
&& &
 (6) & (2) & (0.1) & (2) & (3) &
   & (21) & (4) & (6) & (0.6) & (3) &
 & (5) &
 & (3) &
 & (4) & (1)
\\
${\rm NLO^*}$
&& &
 36 & 1 & 0.1 & 2.9 & 2.4 &
   & 38 & 16 & 4 & 0.3 & 2 &
 & 8.1 &
 & 6 &
 & 8.8 & 2.2
\\
Exp
&& &
 34(16) & $-$ & $-$ & 18(17) & $-$ &
   & 36(22) & 8.5(6.4) & $-$ & $-$ & $-$&
 & 13(5.3) &
 & $-$ &
 & $-$ & $-$
\\[1mm]
\hline \hline
\\
\end{tabular}
\begin{tabular}{cc ccccc c ccccc c ccc} \hline \hline
&&
\multicolumn{5}{c}{$\Delta_{\frac{1}{2}}(1910)$} &
&\multicolumn{5}{c}{$\Delta _{\frac32}(1920)$}  &
&\multicolumn{3}{c}{$\Delta _{\frac{5}{2}}(1905)$ }
\\[1mm] \cline{3-7}\cline{9-13}\cline{15-17}
&
&$\pi  N $&$  K \Sigma  $&$ \pi  \Delta  $&$\eta  \Delta $&$  K \Sigma^* $&$
$&$ \pi  N $&$  K \Sigma  $&$ \pi  \Delta  $&$\eta  \Delta $&$  K \Sigma^* $&$
$&$  \pi  \Delta  $&$\eta  \Delta$&$  K \Sigma^*$
\\[1mm] \cline{3-7}\cline{9-13}\cline{15-17}
LO &
& 31   & 0 & 3.5 & 0 & 0 &
& 17 & 0 & 13  & 0 & 0 &
& 14.5 & 0 & 0
\\
${\rm LO^*}$ &
& 35   & 13 & 6 & 0.4 & 0 &
& 19 & 8.6 & 23  & 2.2 & 0.8 &
& 26 & 1.7 & 0.1
\\
NLO &
&  45 & 9.9 & 5.7 & 1 & 0 &
&  22 & 6.5 & 19 & 5.2 & 0.6 &
&  18 &  4  & 0.1
\vspace{-0.25cm}\\
 &
&  (25) & (6) & (8) & (0.6) & $-$ &
&  (13) & (4) & (15) & (3) & (0.3) &
&  (11) &  (3)  & (0.03)
\\
${\rm NLO^*}$ &
&  41 & 15 & 9 & 0.6 & 0.1 &
&  18 & 8.3 & 28 & 2.5 & 0.9 &
&  22 &  1.5  & 0.1
\\
Exp &
& 52(20) & $-$& $-$& $-$& $-$&
& 28(19) & $-$& $-$& $-$& $-$ &
& $-$& $-$& $-$
\\[1mm] \hline \hline
\end{tabular}
\end{table}
\begin{table}
\parbox{5.5in}{\caption{Fit parameters and partial widths in MeV for F-wave decays of the $[56,2^+]$ baryons.}}\\
\vspace{.5cm}
\begin{tabular}{c c cc c ccccccc}
\hline \hline
     &&  $\chi^2_{dof}$ & dof &&   $C_1$     && $C_2$       && $C_3$       && $B_1$
\\[1mm]
\hline
LO   &&  3.7    & 11   && 0.44(0.03) && $-$         && $-$         && $-$
\\
${\rm LO^*}$   &&  3.4    & 10   &&   0.41(0.02)&& $-$         && $-$         &&  -0.20(0.07)\\
NLO  &&  0.7   & 6   &&  0.52(0.02) &&  -0.76(0.10)   &&  0.42(0.18)  &&  0.26(0.08)
\\
${\rm NLO^*}$  &&  0.83   & 7   &&   0.51(0.02)&& -0.32(0.06)  && 0.40(0.13)  &&  -0.11(0.07)
\\[1mm]
\hline \hline
\\
\end{tabular}
\begin{tabular}{ccc ccccc c ccccc c c c c c cc}
\hline \hline
&&& {$N_{\frac{3}{2}}(1720)$} &&
    {$\Lambda _{\frac{3}{2}}(1890)$}&
   & \multicolumn{3}{c} {$\Delta_{\frac{3}{2}}(1920)$} &
   & \multicolumn{4}{c}{$N_{\frac{5}{2}}(1680)$} &
   & \multicolumn{5}{c}{$\Lambda_{\frac{5}{2}}(1820)$ }
\\[1mm]
\cline{4-4}\cline{6-6}\cline{8-10}\cline{12-15}\cline{17-21}
&&& $\pi \Delta$ &   & $\pi \Sigma^*$ &  & $\pi \Delta$ & $\eta \Delta$ & $K \Sigma^*$ & 
  &  $\pi N$ &  $\pi \Delta$ & $\eta N$ & $K \Lambda$  & 
  & $\bar{K} N$  &$K  \Xi$& $\pi  \Sigma$ & $\eta  \Lambda$ & $\pi \Sigma^*$ \\
\cline{4-4}\cline{6-6}\cline{8-10}\cline{12-15}\cline{17-21}
LO
& & & 4.5 & & 5.6 & & 8.7 & 0 & 0 & & 22 & 1.1
& 0 & 0 & & 0 & 0 & 12 & 0 & 1
\\
${\rm LO^*}$
& & & 8 & & 7 & & 22 & 0.3 & 0 & & 55 & 1.8
& 0.2 & 0.1 & & 42 & 0 & 14 & 0.6 & 1.3
\\
NLO
& & & 27 & & 22 & & 59 & 0.1 & 0 & & 84 & 6
& 0.1 & 0 & & 12 & 0 & 29 & 0.2 & 4
\vspace{-0.25cm}\\
& & & (3) & & (3) & & (13) & (0.04) & $-$ & & (6) & (0.8)
& (0.03) & $-$ & & (3) & $-$ & (2) & (0.1) & (0.6)
\\
${\rm NLO^*}$
& & & 13 & & 12 & & 77 & 0.8 & 0 & & 83 & 2.8
& 0.3 & 0.1 & & 52 & 0 & 21 & 0.7 & 2
\\
Exp
& & & $-$ & & $-$ & & $-$ & $-$ & $-$ & & 88(8) & $-$
& $-$ & $-$ & & 48(7) & $-$ & 9(3) & $-$ & $-$
\\[1mm]
\hline \hline
\\
\end{tabular}
\begin{tabular}{ccc ccccc c ccccc cc } \hline \hline
&&&
\multicolumn{7}{c}{$\Sigma_{\frac{5}{2}}(1915)$} &
&\multicolumn{5}{c}{$\Delta _{\frac52}(1905)$}
\\[1mm] \cline{4-10}\cline{12-16}
& & & $\bar{K} N$ & $\pi \Sigma$ & $\eta \Sigma$ & $K \Xi$ & $\pi \Lambda $ & $\bar{K} \Delta$  
& $\pi \Sigma^*$ & & $\pi N$ & $\eta \Delta$ & $\pi \Delta$ & $K \Sigma$ & $K \Sigma^*$
\\[1mm] \cline{4-10}\cline{12-16}
LO& & & 0 & 18 & 0 & 0 & 15 & 0 & 0.6 & & 13 & 0 & 12 & 0 & 0
\\
${\rm LO^*}$& & & 3.6 & 29 & 1 & 0.5 & 18 & 2 & 1 & & 20 & 0.2 & 30 & 0.7 & 0
\\
NLO& & & 3 & 44 & 0.4 & 0.1 & 36 & 0.4 & 2 & & 45 & 0.1 & 58 & 0.1 & 0
\vspace{-0.25cm}\\
& & & (0.7) & (3) & (0.1) & (0.03) & (3) & (0.1) & (0.4) & & (11) & (0.02) & (7) & (0.03) & $-$
\\
${\rm NLO^*}$& & & 4.4 & 43 & 1.1 & 0.7 & 27 & 2.6 & 1.6 & & 44 & 0.3 & 65 & 1.2 & 0
\\
Exp& & & 12(7) & $-$ & $-$ & $-$ & $-$ & $-$ & $-$ & & 40(13) & $-$ & $-$ & $-$ & $-$
\\[1mm]
\hline \hline
\\
\end{tabular}
\end{table}
\begin{table}
\begin{tabular}{ccc ccccc c ccccc ccc } \hline \hline
&&&
\multicolumn{5}{c}{$\Delta_{\frac{7}{2}}(1950)$} &
&\multicolumn{8}{c}{$\Sigma _{\frac72}(2030)$}
\\[1mm] \cline{4-8}\cline{10-17}
& & & $\pi N$ & $\eta \Delta$ & $\pi \Delta$ & $K \Sigma$ & $K \Sigma^*$ &
& $\bar{K} N$ & $\pi \Lambda$ & $\pi \Sigma$ & $\pi \Sigma^*$ & $\bar{K} \Delta$ & $\eta \Sigma$ & $K \Xi^*$ & $K \Xi$
\\[1mm] \cline{4-8}\cline{10-17}
LO& & & 72 & 0 & 18 & 0 & 0 & 
& 0 & 41 & 14 & 11 & 0 & 0 & 0 & 0
\\
${ \rm LO^*}$& & & 115 & 0.6 & 45 & 6 & 0 & 
& 43 & 50 & 22 & 18 & 16 & 8 & 0 & 1.8
\\
NLO& & & 123 & 0.2 & 48 & 0.9 & 0 & 
& 36 & 35 & 12 & 17 & 3 & 3 & 0 & 0.5
\vspace{-0.25cm}\\
& & & (21) & (0.1) & (9) & (0.2) & $-$ & 
& (8) & (8) & (3) & (4) & (0.8) & (1) & $-$ & (0.1)
\\
${\rm NLO^*}$ & & & 92 & 0.4 & 36 & 4.5 & 0 & 
& 31 & 40 & 18 & 14 & 11 & 8 & 0 & 1.2
\\
Exp& & & 114(25) & $-$ & $-$ & $-$ & $-$ & 
& 35(7) & 35(7) & 13(5) & 18(9) & $-$ & $-$ & $-$ & $-$
\\[1mm] \hline \hline
\end{tabular}
\end{table}
\end{center}

$\bullet$  F-wave decays: All states with $J>1/2$ have F-wave decays. Here, the LO results are problematic as shown by its  $\chi^2_{dof}=3.7$. This is a one-parameter fit to twelve data. The main issue is that the widths for $N(1680)\to \pi N$ and $\Delta(1905)\to \pi N$ resulting from the fit are too small. The situation for these decays at LO is substantially worse than in the case of the analysis in $SU(4)$ \cite{GSdecays}. This is due to the $\Lambda$ and $\Sigma$ decays, that are included in the fits of the present work.  The LO* fit, for which we use the experimental errors, is substantially better.  This is because  not expanding the matrix elements leads to an important enhancement of the rates $N(1680)\to \pi N$  and $\Delta(1905)\to \pi N$ .  We note that these channels are the main source of the large $\chi^2$ at LO.
Clearly, the NLO corrections must be important. The NLO fit has a minor problem of consistency with the bound $\Gamma_{\rm F-wave}(N(1680)\to  \pi\Delta)<2.6$ MeV, 
  violating the bound  by a relatively small amount. The main  problem is that the
decays $\Lambda(1820) \to K N   $ and $\pi\Sigma $ cannot be simultaneously fitted. The NLO fit gives
$\Gamma(\Lambda(1820) \to K N)  <\Gamma( \Lambda(1820) \to\pi \Sigma )$, which is opposite to the empirical ordering.  We find that by eliminating both channels as inputs to the NLO the $\chi^2$ is acceptable.  The discrepancy  is in part  resolved in the fit NLO*, for which the inconsistency remains in the decay $ \Lambda(1820) \to\pi \Sigma $. We have, therefore, carried out the NLO* fit  without that channel.  Since there are no $\Lambda^*$ baryons in the mass proximity of  $\Lambda(1820) $ with which it could mix, the   inconsistency should be   resolved  either by  higher order terms in the $1/N_c$ expansion, which would be indication of poor convergence of the expansion for that channel, or by a  better empirical value of the width, whose measurement dates back to the 1970s. The $SU(3)$ breaking effects are determined by three input widths, and turn out to be within the expectations. 
Note that the experimental errors for the F-wave widths are significantly smaller than for the P-waves, and thus the predictions for the unknown widths should be better than for the P-wave decays.
   
 
\section{Conclusions}
This work implemented the $1/N_c$ expansion for the decays of positive parity ${\bf 56}$-plet baryons, and analyzed the empirically known partial decay widths to ${\cal{O}}(1/N_c)$ and first order in $SU(3)$ symmetry breaking. As it had been well established in previous work within $SU(4)$, the   decays of the $[{\bf 56},0^+]$ baryons are poorly described at LO,  requiring some large NLO corrections by a 2-body operator, which is an indication of the special  nature of this multiplet. The coefficient of the NLO operator is, however, within natural magnitude.  On the other hand,  the  decays of the $[{\bf 56},2^+]$ baryons are well described, showing for both P- and F-wave decays natural size contributions by the NLO operators.  We have noticed that for a considerable number of matrix elements better fits result if they are not expanded in $1/N_c$,  in particular in the case of the the F-waves. For this reason, we included fits  LO* and NLO* to see that effect.

One point to note is that, in general, the empirical widths have rather large error bars, considerably larger than the 10\%, which would allow for an accurate determination of the NLO effects. This impacts on the  predictions one can deduce from the present analysis; in the case  of the   $[{\bf 56},0^+]$  decays, the predictions should be taken with caution, as in this case the $1/N_c$ expansion seems to be poorly converging. In the case of the $[{\bf 56},2^+]$ decays, the predictions for P-wave widths are within errors of the order of 50\%, while  for the F-waves they are   accurate at the 20\% level, upon  having eliminated  the problematic channel  $ \Lambda(1820) \to\pi \Sigma $. Using the elements given in the paper, it is straightforward to make predictions for the widths of the missing states in the multiplets.

\section*{ACKNOWLEDGEMENTS}

This work was supported by DOE Contract No. DE-AC05-06OR23177 under which JSA operates the Thomas Jefferson
National Accelerator Facility,  by the National Science Foundation (USA) through grant  \#~PHY-0555559 (JLG and ChJ)
and by the Subprogram C\'esar Milstein of SECYT (Argentina) (JLG),   by CONICET (Argentina)  grant \# PIP 02368 and by ANPCyT  (Argentina)
grants \# PICT 04-03-25374 and 07-03-00818 (NNS).
JLG   thanks the Grupo de Particulas y Campos, Centro At\'omico Bariloche, and in particular Professor Roberto Trinchero, for the  hospitality extended to him during the completion of this work. 

\section{Appendix: $SU(3)$ Isoscalar Factors}
This appendix gives the   isoscalar factors, which appear in the matrix elements needed in this work, which correspond to the emission of mesons belonging to an octet of $SU(3)$. They are given  for the irreducible representations of $SU(3)$ as needed in this work for generic $N_c$.  The representations are  given in terms of the two labels defining the Young tableu, namely  $(p,q)$, where $p+2 q=N_c$.  The isoscalar factors needed for the decays into the octet of mesons are denoted by:
\beq
\left(\begin{array}{cc}
     (p,q)   & (1,1)       \\
     Y_1~I_1     & Y_2 ~I_2
\end{array}  \right\Arrowvert \left.     \begin{array}{c}
      (p',q')      \\
     Y~~ I
\end{array} \right)_\gamma.
\eeq
Here, $ (p',q')$ is the representation of the final ground state baryon and $(p,q)$ of the initial  excited baryon.
For baryons, the  correspondences between multiplets for generic odd $N_c$ and $N_c=3$ are as follows: 
$(p=1, q=\frac{N_c-1}{2}) \rightarrow   {\bf 8}$, 
 and 
$(p=3, q=\frac{N_c-3}{2})\rightarrow   {\bf 10}$.
Table VI displays these correspondences more explicitly.
 
\begin{center}
\begin{table}[h]
\caption{Representation correspondences for arbitrary odd $N_c$.   }
\vspace{.5cm}
\begin{tabular}{cc|cc|cc}
\hline \hline 
\multicolumn{2}{c|}{{\bf 8} Baryons} & 
\multicolumn{2}{c|}{{\bf 10} Baryons} &
\multicolumn{2}{c}{Mesons} \\
\multicolumn{2}{c|}{$(p,q) = (1,\frac{N_c-1}{2})$}& 
\multicolumn{2}{c|}{$(p,q) = (3,\frac{N_c-3}{2})$}
&\multicolumn{2}{c}{$(p,q) = (1,1)$} \\[2mm]
\hline
State &  $(Y,I)$ &  State  & $(Y,I)$  & State  & $(Y,I)$ \\ [1mm] \hline &&&&&
\\[-3.02mm]
N & $(\frac{N_c}{3},\frac{1}{2})$ & $\Delta$  & $(\frac{N_c}{3},\frac{3}{2})$ & $\pi$ &  $(0,1)$ \\[2mm]
$\Sigma$&  $(\frac{N_c-3}{3},1)$ &  $\Sigma^*$  &$(\frac{N_c-3}{3},1)$  & $\eta$ & $(0,0)$ \\[2mm]
$\Lambda$&  $(\frac{N_c-3}{3},0)$  & $\Xi^*$  &$(\frac{N_c-6}{3},\frac{1}{2})$  & K & $(1,\frac12)$ \\[2mm]
$\Xi$&  $(\frac{N_c-6}{3},\frac{1}{2})$ & $\Omega$  &$(\frac{N_c-9}{3},0)$  & $\bar{K}$  & $(-1,\frac12)$ \\[2mm]
\hline \hline
\end{tabular}
\\[4mm]
\end{table}
\end{center}

In the following, we give the isoscalar factors as needed for the calculations carried out in this work.
\begin{center}
\begin{table}
\parbox{5in}{\caption{Isoscalar factors for  ${\bf 8} \to {\bf 8}$ decays.
The listed
values should be multiplied by $f_1 = \frac{1}{(N_c+3)}$ and
$f_2 = \frac{1}{(N_c+3)}\sqrt{\frac{(N_c-1)}{(N_c+7)}}$
to obtain the actual isoscalar factors for $\gamma=1$ and $\gamma=2$, respectively.}}\\
\vspace{.3cm}
\begin{tabular}{cc cccc c cccc}
\hline \hline
& &\multicolumn{4}{c}{$N$} &
  &\multicolumn{4}{c}{$\Lambda$}
\\ \cline{3-6}\cline{8-11}
& & $ \eta N$ & $ \pi N$ & $K \Sigma $ & $K \Lambda $ &
  & $ \bar{K}N$ & $\pi \Sigma $ & $\eta \Lambda $ & $K \Xi $
\\ \cline{3-6}\cline{8-11}
&&&&&&
\\[-3mm]
$\gamma=1$ & & $N_c$ & 3 & $\sqrt{3(N_c-1)}$ & $\sqrt{3(N_c+3)}$ &
          & $-\sqrt{\frac{3(N_c+3)}{2}}$ & 0 & $N_c-3$ &
                       3$\sqrt{\frac{N_c-1}{2}}$
\\[2mm]
$\gamma=2$ & & 3 & $-(N_c+6)$ & $\frac{N_c+15}{\sqrt{3(N_c-1)}}$ & $- \sqrt{3(N_c+3)}$ &
          & $\sqrt{\frac{3(N_c+3)}{2}}$ &  $-\sqrt{\frac{(N_c+3)^3}{3(N_c-1)}}$ & 6 &
               $\frac{9-N_c}{\sqrt{2(N_c-1)}}$
\\[3mm]
\hline
\end{tabular}
\\[3mm]
\begin{tabular}{cc cc cc cc}
\hline \hline
      & &\multicolumn{5}{c}{$\Sigma$}
\\ \cline{3-7}
       & & $ \bar{K}N$ & $\eta\Sigma $ & $\pi\Sigma $ & $\pi\Lambda $ & $K\Xi $
\\
\cline{3-7} &&&&&&
\\[-3mm]
$\gamma=1$ & & $3\sqrt{\frac{N_c-1}{2}}$ & $N_c-3$ & $2 \sqrt{6}$ &
          0 & $\sqrt{\frac{3(N_c+3)}{2}}$
\\[2mm]
$\gamma=2$ & & $\frac{N_c+15}{\sqrt{2(N_c-1)}}$ &
    $\frac{2(N_c-9)}{N_c-1}$ & -$\sqrt{\frac{2}{3}} \frac{(N_c-3)(N_c+7)}{N_c-1}$ &
       $\sqrt{\frac{(N_c+3)^3}{N_c-1}}$ & $-\frac{5N_c+3}{N_c-1}
       \sqrt{\frac{N_c+3}{6}}$
\\[3mm] \hline
\end{tabular}
\\[3mm]
\begin{tabular}{cc cc cc}
\hline \hline & &\multicolumn{4}{c}{$\Xi$}
\\
\cline{3-6} & & $ \bar{K}\Sigma$ & $ \bar{K}\Lambda$ & $\eta \Xi $ & $\pi \Xi $
\\
\cline{3-6} &&&&&
\\[-3mm]
$\gamma=1$ & & -$\sqrt{N_c+3}$ & 3$\sqrt{N_c-1}$ & $N_c-6$ & 3
\\[2mm]
$\gamma=2$ & & $\frac{5N_c+3}{3(N_c-1)} \sqrt{N_c+3}$ &
$\frac{9-N_c}{\sqrt{N_c-1}}$ & $\frac{7N_c-15}{N_c-1}$ &
$\frac{N_c^2 + 3 N_c +36}{3(N_c-1)}$
\\[2mm] \hline
\end{tabular}
\end{table}
\end{center}

\begin{center}
\begin{table}
\parbox{5in}{\caption{Isoscalar factors for   ${\bf 10} \to {\bf 10}$  decays. The listed
values should be multiplied by $f_1 = \frac{1}{\sqrt{45+N_c(N_c+6)}}$ and
$f_2 = \sqrt{\frac{5(N_c-3)(N_c+5)}{(N_c+1)(N_c+9)(45+N_c(N_c+6))}}$
to obtain the actual isoscalar factors for $\gamma=1$ and $\gamma=2$, respectively.
}}\\ \vspace{.3cm}
\begin{tabular}{cc ccc c cccc}
\hline \hline
& & \multicolumn{3}{c}{$\Delta$} &
  & \multicolumn{4}{c}{$\Sigma^*$}
\\
\cline{3-5}\cline{7-10}
& & $\eta \Delta $ & $\pi \Delta $ & $K \Sigma^* $ &
  & $ \bar{K}\Delta$ & $\eta \Sigma^* $ & $\pi \Sigma^* $ & $K \Xi^* $
\\
\cline{3-5}\cline{7-10}
&&&&& &&&&
\\[-3mm]
$\gamma=1$& & $N_c$ & $3 \sqrt{5}$ & $\sqrt{3(N_c+5)}$ &
         & $-\frac{3 \sqrt{N_c+5}}{2}$ & $N_c-3$ & $2 \sqrt{6}$ & $\sqrt{6(N_c+3)}$
\\[2mm]
$\gamma=2$& & 3 & $-\frac{N_c+6}{\sqrt{5}}$ & $\frac{3-N_c}{\sqrt{3(N_c+5)}}$ &
         & $\frac{N_c-3}{2 \sqrt{N_c+5}} $ & $\frac{4(N_c+3)}{N_c+5}$ &
                   $-\frac{N_c^2 + 10 N_c + 33}{\sqrt{6}
                  (N_c+5)}$ & $\frac{3-N_c}{N_c+5} \sqrt{\frac{2(N_c+3)}{3}}$
\\[2mm] \hline
\end{tabular}
\\[3mm]
\begin{tabular}{c c cc c c ccc}
\hline \hline
& &\multicolumn{4}{c}{$\Xi^*$} &
  &\multicolumn{2}{c}{$\Omega$}
\\
\cline{3-6} \cline{8-9}
& & $ \bar{K}\Sigma^*$ & $\eta\Xi^* $ & $\pi \Xi^* $ & $K \Omega $ &
  & $ \bar{K} \Xi^*$ & $\eta \Omega $
\\
\cline{3-6} \cline{8-9} & & & & &
\\[-3mm]
$\gamma=1$& &$-2 \sqrt{N_c+3}$ & $N_c-6$ & 3 & $3 \sqrt{N_c+1}$ &
         & $-3 \sqrt{\frac{N_c+1}{2}}$ & $N_c-9$
\\[2mm]
$\gamma=2$& & $\frac{2(N_c-3) \sqrt{N_c+3}}{3 (N_c+5)} $ &
           $\frac{5N_c+9}{N_c+5}$ & $-\frac{N_c^2+ 9 N_c+36}{3(N_c+5)}$ &
           $\frac{(3-N_c) \sqrt{N_c+1}}{N_c+5}$ &
         & $\frac{N_c-3}{N_c+5} \sqrt{\frac{N_c+1}{2}}$ & $\frac{6(N_c+1)}{N_c+5}$
\\[2mm] \hline
\\
\end{tabular}
\end{table}
\end{center}

\begin{center}
\begin{table}
\parbox{5in}{\caption{Isoscalar factors for    ${\bf 8} \to {\bf 10}$ decays. The listed
values should be multiplied by
$f= \frac{\sqrt{2}}{\sqrt{(N_c+1)(N_c+5)}}$ to obtain
the actual isoscalar factors.}}\\
\vspace{.3cm}
\begin{tabular}{ccc ccc c}
\hline \hline
& &\multicolumn{2}{c}{$N$}&
  &\multicolumn{2}{c}{$\Lambda$}
\\
\cline{3-4}\cline{6-7}
& & $\ \ \ \ \ \ \pi \Delta \ \ \ \ \ \ $ & $\ \ \ \ \ \ K \Sigma^* \ \ \ \ \ \ $ &
  & $\ \ \ \ \ \ \ \ \ \pi  \Sigma^* \ \ \ \ \ \ \ \ \ $ & $\ \ \ \ \ \ K \Xi^* \ \ \ \ \ \ $
\\
\cline{3-4}\cline{6-7}
&&& &&&
\\[-3mm]
& & -$\sqrt{\frac{(N_c-1)(N_c+5)}{2}}$ & -$2\sqrt{\frac{N_c-1}{3}}$ &
  & $-\sqrt{\frac{(N_c+3)(N_c-1)}{3}}$ &  $-\sqrt{2 (N_c-1)}$
\\[2mm] \hline
\end{tabular}
\\[3mm]
\begin{tabular}{cc cccc c cccc}
\hline \hline
& &\multicolumn{4}{c}{$\Sigma$} &
  & \multicolumn{4}{c}{$\Xi$}
\\
\cline{3-6} \cline{8-11}
& & $\ \ \   \bar{K}\Delta \ \ \ $ & $\ \ \  \eta \Sigma^*  \ \ \ $ & $\ \ \  \pi \Sigma^* \ \ \ $ & $\ \ \ K \Xi^*  \ \ \ $ &
  & $\ \ \  \eta \Xi^* \ \ \ $ & $\ \ \  \pi \Xi^*\ \ \ $ & $\ \ \  \bar{K} \Sigma^* \ \ \ $ & $\ \ \  K \Omega \ \ \ $
\\
\cline{3-6} \cline{8-11}
&&&&& &&&&&
\\[-3mm]
& & $\sqrt{N_c+5}$ & 2 & $\frac{N_c+1}{\sqrt{6}}$ & $\sqrt{\frac{2(N_c+3)}{3}}$ &
  & 2 & $\frac{2 N_c}{3}$ & $\frac{2\sqrt{N_c+3}}{3}$ & $2 \sqrt{N_c+1}$
\\[2mm] \hline
\\
\end{tabular}
\end{table}
\end{center}


\begin{center}
\begin{table}
\parbox{5in}{\caption{Isoscalar factors for    ${\bf 10} \to {\bf 8}$ decays. The listed
values should be multiplied by
$f=\left( (N_c+7)(N_c-1) \right)^{-1/2}$
to obtain the actual isoscalar factors.}}\\
\vspace{.3cm}
\begin{tabular}{cc cc cc cc cc }
\hline \hline & &\multicolumn{2}{c}{$\Delta$}& & &
\multicolumn{4}{c}{$\Xi^*$}
\\
\cline{3-4}\cline{7-10}
 & & $\pi N $ & $K \Sigma $ & & & $\ \ \ \  \bar{K} \Sigma\ \ \ \ $ & $\ \   \bar{K}\Lambda  \ \ $ & $\ \  \eta\Xi   \ \ $ & $\ \  \pi \Xi \ \ $
\\
\cline{3-4}\cline{7-10} & & & & & & & &
\\[-3mm]
& & $-\sqrt{(N_c-1)(N_c+5)}$ & $2 \sqrt{\frac{N_c+5}{3}}$ & &
  & $\frac{2 \sqrt{N_c+3}}{3}$ & $2 \sqrt{N_c-1}$ & $-2$ & $-\frac{2N_c}{3}$
\\[2mm] \hline
\end{tabular}
\\[3mm]
\begin{tabular}{cc cc cc cc cc }
\hline \hline & &\multicolumn{5}{c}{$\Sigma^*$}& & & {$\Omega$}
\\
\cline{3-7}\cline{10-10}
 & & $\  \eta \Sigma \ $ & $ \bar{K} N $ & $\ \ \ \ \pi  \Sigma  \ \ \ \ $ & $\pi\Lambda $ & $K\Xi $ & & & $ \bar{K}\Xi $
\\
\cline{3-7}\cline{10-10}& & & & & & & & &
\\[-3mm]
 & & $-2$ & $\sqrt{2(N_c-1)}$ & $-\frac{N_c+1}{\sqrt{6}}$ & $-\sqrt{(N_c+3)(N_c-1)}$ & $\sqrt{\frac{2(N_c+3)}{3}}$
& & & $\sqrt{2(N_c+1)}$
\\[2mm] \hline
\\
\end{tabular}
\end{table}
\end{center}

\pagebreak

\end{document}